\documentclass{elsart}

\usepackage{natbib}

\usepackage{graphicx}

\usepackage{amssymb}

\def\arcsec{\hbox{$^{\prime\prime}$}}

\begin{document}

\begin{frontmatter}



\title{X-ray synchrotron emission from supernova remnants}


\author{Jean Ballet}
\ead{jballet@cea.fr}
\address{DSM/DAPNIA/SAp, Bt 709, CEA Saclay, 91191 Gif sur Yvette Cedex, 
France}

\begin{abstract}
X-ray synchrotron emission tells us of the highest energy reached by 
accelerated electrons. In a few supernova remnants (SN 1006, G347.3$-$0.5)
this is the dominant form of X-ray radiation, but in most it is superposed to 
the dominant thermal emission.
Thanks to the spectro-imaging capability of 
{\sl Chandra} and {\sl XMM-Newton}, X-ray 
synchrotron emission has now been unambiguously detected in most young
supernova remnants 
(Cas A, Tycho, Kepler). It arises in a very thin shell (a few arcsecs) at 
the blast wave. The thinness of that shell (much broader in the radio domain) 
implies that the high energy electrons cool down very fast behind the shock. 
The magnetic field that one deduces from that constraint is more than
100 $\mu$G behind the shock. 
\end{abstract}

\begin{keyword}
Acceleration of particles \sep Magnetic fields \sep
Cosmic rays \sep ISM: supernova remnants \sep
X-rays
\PACS 98.38.Mz \sep 98.70.Sa
\end{keyword}

\end{frontmatter}

\section{Introduction}

The current paradigm is that the bulk of the cosmic-rays 
(up to the ``knee'' at $3 \times 10^{15}$ eV) 
are accelerated at the blast waves
generated by supernova explosions in our galaxy \citep{be87}.
Indeed supernova remnants (SNRs) are all non-thermal radio emitters, attesting
of the presence of accelerated electrons at energies of 1 GeV or so
in larger amounts than in the average interstellar medium.
In most SNRs, the radio emission is limb-brightened, confirming that those 
accelerated electrons originate at the shock rather than in a central pulsar.

The X-ray emission in those shell-type SNRs
is usually mostly thermal (dominated by strong lines
of heavy elements), but the synchrotron emission may extend up to
the X-rays and contribute to the emission as well.
In several cases, like SN 1006 \citep{ko95} and G347.3$-$0.5 \citep{kk97},
the synchrotron component actually dominates the X-ray emission.

The slope of the X-ray synchrotron emission is always much steeper
($\alpha \simeq 1.5$)
than that of the radio emission ($\alpha \simeq 0.5$).
This indicates that the X-rays (emitted by electrons at energies $>$ 1 TeV)
probe the cut-off of the electron distribution, and can
tell something about the limits of the shock acceleration mechanism.
The open questions on which observations of X-ray synchrotron
in SNRs may shed light are:
\vspace{-\medskipamount}
\begin{itemize}
\item How efficient is cosmic-ray acceleration in SNRs ? 
      What is the energy density of accelerated particles ?
\item What is the maximum energy of accelerated particles ?
\item How large is the magnetic field ? Is it very turbulent ?
      Is it amplified ?
\end{itemize}

Unfortunately, not much is known about the synchrotron emission
in the seven decades gap between the radio and the X-rays, partly
because it is masked by other components (like infra-red emission
by heated dust) and partly because it is simply very difficult
to detect faint extended emission.
Because of that ignorance, it is often assumed that the spectrum
extends as a power law up to the cut-off, even though non linear
acceleration models predict some concavity \citep{bae99}.
The synchrotron emission has been detected in Cas A in the near infra-red
\citep{jr03}, but this is associated with the bulk of the radio
emission (near the interface with the ejecta), not with the relatively
faint synchrotron emission from behind the blast wave discussed
in the following sections.

In this review I report on recent observations of X-ray synchrotron
emission in three young SNRs dominated by thermal emission
from the ejecta (Cas A, Kepler and Tycho), and two older SNRs dominated
by non-thermal emission (SN 1006 and G347.3$-$0.5).
The inference is that the magnetic field just behind the
blast wave is quite large (up to 200 $\mu$G).

\section{Observations at high angular resolution}

Before it could be spatially resolved, the X-ray emission behind the blast wave
in young SNRs was expected to be the thermal emission of 
the shocked ambient gas,
distributed more or less evenly between the blast wave 
and the outer boundary of the ejecta.
But the {\sl Chandra} results challenged that preconception.
The images of the continuum emission (4 to 6 keV) in Cas A \citep{gk01}
and Tycho \citep{hd02} clearly
show a very thin X-ray rim at
the blast wave.
The apparent width of the rim is less than 4$\arcsec$ or 
2 10$^{17}$ cm in Cas A, 
and appears to be twice lower at places.
That X-ray rim runs all around both remnants, albeit in a broken manner
(particularly in Cas A).

The 4 to 6 keV continuum emission in Kepler shows a very similar structure.
Sharp (although not nearly as sharp) rims have also been observed
in SN 1006 \citep{lr03, by03} and G347.3$-$0.5 \citep{ls03}.
In those two SNRs, older and dominated by non-thermal emission, and
contrary to the younger three mentioned above, there
is no clear boundary behind which the thermal emission 
from the ejecta dominates. 
In G347.3$-$0.5, the thermal emission is not even detected at all.
Also, because the non-thermal emission dominates,
the energy range where its morphology can be studied is not restricted to
the 4 to 6 keV band (SN 1006 is very faint in that band).

In Tycho, the rim clearly marks the outer boundary of the X-ray emission, 
so it is natural to think of it as a thin sheet covering
the entire sphere, rather than a linear filament.
This is supported by the observation of SN 1006, where the rim
is definitely resolved, and much sharper outwards than inwards.
The visual appearance of the rims in Cas A is much less regular.
This is not a strong argument against the sheet geometry, however, because
it is actually very reminiscent of the optical emission in older remnants
like the Cygnus Loop, which is generally interpreted as due to
a wrinkled sheet of emission \citep{he87}.
In keeping with this evidence, I will assume the rims are places 
where the sheet
of emission at the blast wave is observed tangentially.

As noted by \citet{bv04}, 
since this is observed in projection the scale height
of the spherical layer must be even smaller than the observed width.
I argue in App.~\ref{project} that it should be 4.6 times smaller.
This conclusion does not depend on the curvature radius of the sheet
(it does not have to be equal to the SNR radius, it could even
be negative -- outwards), therefore it is quite robust.
Exceptional geometrical conditions \citep[like model B of][]{he87},
normally associate a narrower rim with a larger brightness contrast 
between the rim
and its surroundings. This should be testable statistically.

The brightness contrast of the X-ray continuum between at the rim
and behind the rim is very large (typically 5). It is consistent with
the thin sheet model in Cas A \citep{bv04} and Tycho \citep{vb05}, 
but leaves little room for anything else.
This implies that most of the volume between the blast wave and the interface
with the ejecta is actually X-ray dark (in the continuum emission). 

Another important observation is the nearly featureless
nature of the spectrum (very faint lines are observed). 
In a thermal framework, this can be explained
if ionization is very far out of equilibrium \citep[][on Tycho]{hd02}.
But of course it is very natural if the spectrum of the rims
is non-thermal. It is important to note in that respect that
the line emission is not peaked at the rims.
The brightness in the lines is essentially the same in a region
behind the rims as on the rims in Cas A and Tycho.
This suggests that the faint line emission comes from the entire volume
between the blast wave and the bright ejecta, 
contrary to most of the continuum emission.

\section{The nature of the X-ray emission behind the blast wave}

The observed geometry is inconsistent with thermal models 
in a uniform medium which predict
emission everywhere up to the interface, with only a slight maximum
at the blast wave. This region is full of hot gas, which cannot cool down
efficiently at those low densities, so it has to be entirely X-ray bright.
Such a sharp decline of the X-ray emission behind the blast wave
could be due to a recent density increase of the ambient gas, but it would
have to be an extraordinary coincidence that this happens exactly
at the same time all around the remnants in both Cas A and Tycho.
In Cas A (a type II SN), having just hit a spherical wind shell could 
create such a condition, but it would result in a much slower shock, 
contrary to the observed proper motions
 \citep{dr03}.

Other known transient phenomena related to the thermal gas 
cannot explain the observations at all:
\vspace{-\medskipamount}
\begin{itemize}
\item Heavy element ionization affects the lines rather
      than the continuum.
\item Electron heating by the ions predicts a hardening
      of the spectrum, but not a decrease in intensity.
\item Dust destruction results in increasing emission.
\end{itemize}

The only other possible source of X-ray radiation is the accelerated particles.
Non-thermal bremsstrahlung (by suprathermal electrons at 10 keV or so)
could be an option.
The density of targets (the thermal gas) does not decline steeply behind
the shock, but collisional losses with the thermal electrons could be 
strong enough to get rid of the particles themselves as they are advected
downstream.
A gross estimate of the energy loss time for suprathermal electrons 
at energy $E_{\mathrm{keV}}$ in gas of electronic density $n_e$ is
$t_{\mathrm{Coul}} = 4.9 \times 10^7$ s $E_{\mathrm{keV}}^{3/2} / n_e$.
A more accurate formula may be found in \citet{ss97}.
Around 5 keV, this could be consistent with the observed width in Cas A
(corresponding to an age of 2.5 $\times 10^8$ s $r/4$, where $r$ is
the total compression ratio) for a downstream gas density
of 1.8 cm$^{-3} (r/4)^{-1}$.
However, non-thermal bremsstrahlung must be associated with brighter 
thermal emission. 
In Cas A, Tycho or Kepler, no such bright thermal emission is observed
associated with the X-ray rims.
With an interstellar column density $N_{\mathrm H}$ 
of several 10$^{21}$ cm$^{-2}$,
it is conceivable that the thermal gas could hide at low temperature
around 1 or 2 $\times 10^6$ K. Such a low temperature is actually predicted
by strongly non linear acceleration models \citep{de00}.
SN 1006, on the other hand, 
has much lower $N_{\mathrm H} \simeq 7 \times 10^{20}$ cm$^{-2}$
\citep{dg02}. The soft thermal emission (in particular the O K lines) 
is clearly seen there at the same
place as the harder X-ray rims.
However it is not brighter than
the non-thermal emission at all, which eliminates the possibility
that the non-thermal continuum could be a tail of the thermal emission.

The most natural remaining option is synchrotron emission 
by high energy electrons.
Here again, advection of the particles and the magnetic field (with only
slight adiabatic losses) cannot explain the very sharp drop behind the shock.
On the other hand, the particles may lose their energy radiatively fast enough
as they are advected (so that their synchrotron emission is shifted below the
X-ray range) to explain a very thin emission region if the magnetic
field is large enough \citep{bam03,vl03}. In Sect. \ref{numbers}
I develop the consequences of that interpretation.

The synchrotron interpretation implies also that
the density (or the temperature) of the ambient gas must be low enough
that it does not contribute significantly to the X-ray emission. 
This is possible
if $n$ (shocked) $<$ 0.15 cm$^{-3}$ in Kepler \citep{cdk04}.
The constraint for Tycho is probably a little less severe.
Another prediction is that the radio emission should be much broader
(not so peaked at the rims), because the GeV electrons emitting
the radio are not affected by losses.
This is indeed what is observed in Cas A \citep{vl03}.
In SN 1006 \citep[][Fig. 4]{lr03}, it is true as well that the non-thermal 
X-ray emission (above 1.2 keV) is much sharper than the radio emission.
The radio is never more peaked than the X-ray emission below 0.8 keV,
which is mostly thermal (O K lines).
In Tycho on the other hand, a radio peak just behind the blast wave 
(associated with the X-ray rim) seems to exist \citep{db91}. This is difficult
to understand in the synchrotron interpretation of the X-rays.

Recently, \citet{lp04} suggested that the narrow rims could mark
the interface between the ejecta and the shocked ambient gas,
rather than the blast wave. Indeed it is relatively natural
to reach high magnetic fields there.
However, I do not think this interpretation is tenable for three reasons:
\vspace{-\medskipamount}
\begin{itemize}
\item Optical H$\alpha$ observations show filaments at the same distance
      from the SNR center as the X-ray rims (although not coincident
      with them). The line shape indicates that a fast (originally ionized)
      and a slow (originally neutral) population exist there. It is hard
      to imagine what could cause this far from a shock.
\item No strong metal lines are observed immediately behind the rims,
      even in SN Ia (Tycho), where no hydrogen envelope should exist
      just behind the interface.
\item One would have to find another explanation for the bulk of
      the radio emission (far behind the X-ray rims), which is currently
      interpreted as originating in the high magnetic fields at the interface.
\end{itemize}

\section{Physical conditions behind the blast wave}
\label{numbers}

Quantitatively, the interpretation of the X-ray rims as synchrotron goes
as follows. The synchrotron cooling time depends on the magnetic
field and the electron's energy as
$t_{\mathrm{cool}} = 6.37 \times 10^8 \; B_{\mathrm{mG}}^{-2} \;
                     E_{\mathrm{erg}}^{-1} \;$ s.
Since the characteristic frequency at which an electron radiates is
$\nu_{\mathrm{sync}} = 1.82 \times 10^{15} \; B_{\mathrm{mG}} \;
                       E_{\mathrm{erg}}^2 \;$ Hz, 
the cooling time may be expressed as a function
of the frequency at which the rims are observed.
\begin{equation}
t_{\mathrm{cool}} = 5.5 \times 10^7 \; B_{\mathrm{mG}}^{-3/2} \;
                    \nu_{\mathrm{keV}}^{-1/2} \;\; \mathrm s
\end{equation}

Two effects (expected to be of the same order at the maximum energy
reached by loss-limited electrons) combine to set the scale height
of the emission behind the shock: advection \citep{bam03,vl03,by03}
and diffusion \citep{bk03,yy04}.
\vspace{-\medskipamount}
\begin{eqnarray}
l_{\mathrm{adv}} & \; = \; & t_{\mathrm{cool}} v_{\mathrm{sh}} / r \; = \;
   1.8 \times 10^3 \; B_{\mathrm{mG}}^{-3/2} \; \nu_{\mathrm{keV}}^{-1/2}
   \; v_{1000} / r \;\; \mathrm{pc} \\
l_{\mathrm{dif}} & \; = \; & \sqrt{\kappa_{\mathrm d} t_{\mathrm{cool}}} \; = 
   \; 1.2 \times 10^{-3} \; B_{\mathrm{mG}}^{-3/2} \;\; \mathrm{pc}
\label{ldif} \\
l_{\mathrm{adv}} / l_{\mathrm{dif}} & \; = \; & 1.5 \;
   \nu_{\mathrm{keV}}^{-1/2} \; v_{1000} / r
\label{ratio}
\end{eqnarray}
where $v_{\mathrm{sh}}$ is the shock speed,
$v_{1000} = v_{\mathrm{sh}}$/(1000 km/s), $r$ is the total compression ratio
and $\kappa_{\mathrm d} = c/(3e) \, E/B$ is the downstream 
diffusion coefficient in the Bohm limit.

\begin{table}
\begin{tabular}{c|c|c|c|c}
Name & Distance & Shock speed & Age & Cut-off frequency \\
\hline
Cas A & 3.4 kpc $^1$ & 5200 km/s $^2$ & 320 yr ? $^4$ & 1.2 keV $^3$  \\
Kepler & 4.8 kpc $^5$ & 5400 km/s $^6$ & 400 yr & 0.9 keV $^8$  \\
Tycho & 2.3 kpc $^9$ & 4600 km/s $^{10}$ & 430 yr & 0.29 keV $^{11}$  \\
SN 1006 & 2.2 kpc $^{12}$ & 2900 km/s $^{13}$ & 1000 yr & 3 keV $^{15}$  \\
G347.3$-$0.5 & 1.3 kpc $^{16}$ & 4000 km/s ? $^{17}$ & 1620 yr ? $^{19}$ & 2.6 keV $^{18}$ \\
\hline
\hline
Name & Obs. freq. & $l_{\mathrm{adv}}/l_{\mathrm{dif}}$ & Projected width & Magnetic field \\
\hline
Cas A & 5 keV & 0.88 & 0.05 pc (3\arcsec) $^3$ & 230 $\mu$G \\
Kepler & 5 keV & 0.92 & 0.07 pc (3\arcsec) $^7$ & 180 $\mu$G \\
Tycho & 5 keV & 0.78 & 0.05 pc (4\arcsec) $^{11}$ & 250 $\mu$G \\
SN 1006 & 2 keV & 0.78 & 0.2 pc (20\arcsec) $^{14}$ & 87 $\mu$G \\
G347.3$-$0.5 & 2 keV & 1.06 & 0.25 pc (40\arcsec) $^{18}$ & 79 $\mu$G \\
\end{tabular}
\smallskip

\caption{Characteristics of the non-thermal emission behind the blast wave
in young SNRs. The cut-off frequency is rather uncertain.
The magnetic field is always estimated from equating $l_{\mathrm{dif}}$ 
(Eq. \ref{ldif}) and the projected width divided by 4.6 
(App. \ref{project}), even
when $l_{\mathrm{adv}}/l_{\mathrm{dif}}$ (from Eq. \ref{ratio})
is formally larger than 1.\newline
References: $^1$ \citet{rh95}, $^2$ \citet{vb98}, 
$^3$ \citet{vl03} from a fit to the spectrum of the whole SNR, 
$^4$ \citet{as80}, $^5$ \citet{rg99}, $^6$ \citet{hu99}+distance, 
$^7$ Decourchelle (private communication), 
$^8$ \citet{cdk04} in the southeast, 
$^9$ \citet{sk91}, $^{10}$ \citet{hu00},
$^{11}$ \citet{hd02}, $^{12}$ \citet{wg03},
$^{13}$ \citet{gw02} in the northwest, 
$^{14}$ \citet{by03} in the northeast, $^{15}$ \citet{rb04} in the northeast,
$^{16}$ \citet{cdg04}, $^{17}$ from $v = \lambda r/t$ taking 
the expansion parameter $\lambda = 2/3$, 
$^{18}$ \citet{ls03} in the northwest, $^{19}$ \citet{wq97}
\smallskip
}
\label{nt_snrs}
\end{table}

The observed characteristics of the five young shell-type SNRs
for which enough data exists are summarized in Table \ref{nt_snrs},
together with the required magnetic field.
The ratio $l_{\mathrm{adv}}/l_{\mathrm{dif}}$ is always estimated
assuming a compression ratio of 4. If the compression ratio is larger
(as predicted by non-linear acceleration models), then $l_{\mathrm{adv}}$
will be smaller but the magnetic field estimate 
(from $l_{\mathrm{dif}}$) won't change.

Overall, this interpretation imposes B $\simeq$ 200 $\mu$G downstream
in the three youngest SNRs.
It is a very important result because it provides observational
evidence for the idea \citep{blu01} that diffusively accelerated particles
streaming ahead of the shock are able to generate a turbulent magnetic
field larger than the original ordered field (which cannot be larger
than a few $\mu$G in such surroundings). This is potentially the key for
breaking the \citet{lc83} limit and
accelerating protons and heavier ions (not limited by radiative losses)
up to the 'knee' 
of the cosmic-ray distribution at 3 10$^{15}$ eV.
An observational consequence is that it predicts weak
inverse Compton emission in the TeV range, because the density
of accelerated electrons (for a given synchrotron emission)
is much lower than estimated from a 'standard' compressed field 
of 10 $\mu$G or so.

A somewhat puzzling aspect is the geometry of the acceleration
with respect to the magnetic field.
In G347.3$-$0.5 and Kepler, the very large asymmetry in the X-ray image
is probably due to variations of the exterior conditions (ambient density),
so it is difficult to infer anything on the geometry.
In SN 1006, the non-thermal emission is very clearly bipolar.
The very faint X-ray emission from the center of the remnant
above 2 keV led \citet{rb04} to conclude that the non-thermal X-rays
are emitted in two polar caps.
This is coherent with theoretical ideas \citep{vb03} that acceleration
works better where the magnetic field was originally parallel
(to the shock velocity).
Unfortunately, that nice picture is not at all what we see in Cas A 
\citep{hl04} and Tycho \citep{hd02} where the non-thermal X-ray rims
seem to run all around the remnants.
The main differences between those two and SN 1006 are the shock velocity
and the ambient density (both lower in SN 1006).
It may be that somehow the faster acceleration at larger shock velocities
can work whatever the magnetic orientation, but a quantitative
explanation does not exist yet.

\section{The shock precursor in the ambient gas}

The principle of diffusive acceleration predicts that the accelerated
particles and the magnetic field should also be present some distance
ahead of the shock.
This means that synchrotron emission should be observed at some level
upstream of the blast wave.

The absence of such a precursor in the radio range 
(an upper limit on its width) implies an upper limit on the diffusion
coefficient at the energy of electrons which radiate in the radio.
This is equivalent to a lower limit on the turbulent field
which is larger than the ordinary interstellar turbulence.
This argument led \citet{ab94}
to conclude that accelerated particles could indeed generate the turbulent
field which is required for acceleration to be a fast process.

Because the diffusion coefficient (for a given level of turbulence)
increases with energy, X-ray observations (corresponding to electron energies
10$^4$ times larger than radio observations) provide a much more stringent
constraint.
In Cas A and Tycho, the radial profile of the X-ray emission appears symmetric,
so one might think that the width used above may be used as an upper
limit to the size of the precursor.

It is not so, though, because the shock compression necessarily results
in compression of the magnetic field. Assuming isotropic magnetic
turbulence upstream and shock compression by a factor $r_{\mathrm{sub}}$, 
the magnetic field downstream may be larger than upstream 
by a factor $r_{\mathrm B} = \sqrt{(1+2r_{\mathrm{sub}}^2)/3}$,
and the synchrotron emission by approximately $r_{\mathrm B}^\Gamma$
(because increasing
$B$ moves the spectrum both up and to larger frequencies),
where $\Gamma$ is the X-ray photon index (typically 2.5 to 3).
$r_{\mathrm{sub}}$ here is the compression ratio at the gas subshock.
It is always lower than 4, and decreases when the total compression ratio 
$r$ increases.

In SN 1006, there is no evidence of a precursor either.
\citet{lr03} give a very stringent upper limit of 1.5\% of the postshock level
to the X-ray brightness about 20\arcsec (or 0.2 pc) upstream.
They argue that this could be explained 
with a large magnetic field jump (larger than 4), 
possibly due to magnetic field amplification at the shock.
\citet{by03} argue that it can be due to a perpendicular
(to the shock normal) magnetic field.
But \citet{rb04}, on the basis of the overall
geometry of the X-ray emission, show that this cannot be true
(the bright limbs must be polar caps rather than an equatorial belt).
\citet{bk03} show that the observed profiles are compatible
with the expected emissivity jump for a turbulent field 
($r_{\mathrm B}^\Gamma$ = 12.5 with $r_{\mathrm{sub}}$ = 3.6 
and $\Gamma$ = 2.3), 
plus the intensity decrease outwards on the diffusive scale height.
In other words, this constrains the upstream diffusive scale height
for electrons emitting 1.5 keV X-rays to be lower than 0.12 pc or so.
Assuming Bohm diffusion, the upstream scale height 
$\kappa_{\mathrm u} / v_{\mathrm{sh}}$ deduced from Table
\ref{nt_snrs}, dividing $B$ by $r_{\mathrm B}$ and deducing $E$
corresponding to emission at 2 keV, is 0.08 pc.
This is close enough that the precursor should be detectable
in the near future.

\ack
I acknowledge constant discussion with A. Decourchelle, and many
discussions on particle acceleration with D. Ellison, E. Parizot,
Y. Gallant and A. Marcowith. I am also indebted to the referees
who helped me understand what was not obvious in the picture I present.

\appendix

\section{Projection of a thin sheet along the line of sight}
\label{project}

Let $r$ be the distance to the center of a sphere 
in units of the sphere's radius $R_s$, and $\rho$ the distance to the center
of the disk it projects onto a plane (in the same units).
The projection of a spherically symmetric emissivity
$\mathcal{E}(r)$ onto that plane results in an axisymmetric
brightness profile
\vspace{-\medskipamount}
\begin{eqnarray}
\mathcal{B}(\rho) & = & 2 \, R_s \int_0^{\sqrt{1-\rho^2}} \mathcal{E}(r)
                        \mathrm dz
\label{projfull} \\
r^2 & = & \rho^2 + z^2
\label{pyth}
\end{eqnarray}
where $z$ is the coordinate perpendicular to the plane.

\begin{figure}
   \centering
   \includegraphics[width=\columnwidth]{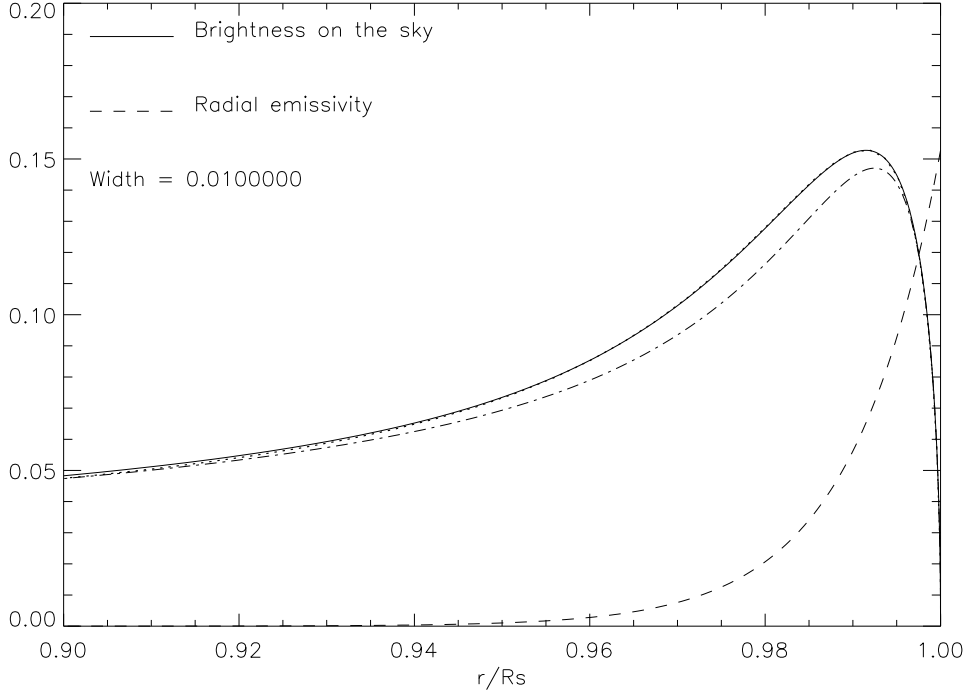}
   \caption{Emissivity profile exponentially decreasing inward
            (Eq.~\ref{expon} with $a = 0.01$, dashed line)
            compared with brightness profile
            projected onto the sky (Eq.~\ref{projfull}, solid line).
            The brightness units are $R_s \times$ the emissivity at the shock,
            where $R_s$ is the shock radius.
            On that scale, the brightness at the center of the disk is 0.02.
            The emissivity profile was scaled down (for illustration purposes)
            to the same maximum.
            The dotted line (barely distinguishable from the solid one)
            is the result of the thin sheet approximation (Eq.~\ref{projthin}).
            The dash-dotted line is Eq. 6 of \citet{bv04}.
           }
   \label{profiles}
\end{figure}

If the emissivity is concentrated near $r = 1$, it may be written
$\mathcal{E}(r) = f(x)$ where $x = (1-r)/a$, $f$ is the functional form
of the decrease and $a$ is the scale height (in units of $R_s$).
Defining $y = (1-\rho)/a$ and $u = z / \sqrt{2a}$,
$r$ may be developed for $\rho$ close to 1
at order 2 in powers of $z$ to give
\vspace{-\medskipamount}
\begin{eqnarray}
r & \simeq & \rho \left( 1 + \frac{z^2}{2\rho^2} \right) \\
\frac{1-r}{a} & \simeq & y - u^2 \\
\mathcal{B}(\rho) & \simeq & 2 \, R_s \, \sqrt{2a}
                        \int_0^{\sqrt{y}} f(y-u^2) \mathrm du
\label{projthin}
\end{eqnarray}
Eq.~\ref{projthin} is of the form 
$\mathcal{B}(\rho) = 2 R_s \sqrt{2a} \, g(y)$.
It always has the same functional form in terms of $y$ near the edge
of the disk (whatever the width $a$).
In the same limit of small $a$, the brightness toward the center 
of the sphere is
\vspace{-\medskipamount}
\begin{eqnarray}
\mathcal{B}(0) & = & 2 \, R_s \int_0^1 \mathcal{E}(r) \mathrm dr \\
\mathcal{B}(0) & \simeq & 2 \, R_s \, a \int_0^{\infty} f(x) \mathrm dx
\label{projcen}
\end{eqnarray}

The formulae above may be quantified in the special case of exponential
decrease, shown on Fig.~\ref{profiles}.
\vspace{-\medskipamount}
\begin{eqnarray}
\mathcal{E}(r) & = & \exp\left(\frac{r-1}{a}\right)
\label{expon} \\
\mathcal{B}(\rho) & \simeq & 2 \, R_s \, \sqrt{2a} \; e^{-y}
                        \int_0^{\sqrt{y}} \exp\left(u^2\right) \mathrm du
\label{expproj}
\end{eqnarray}
Then the maximum of $g(y)$ occurs at $y_0 = 0.854$ 
where it reaches $g_0 = 0.541$.
$g(y)$ decreases inwards to half that value at $y_1 = 4.685$.
$g(y)$ decreases outwards to half that value at $y_2 = 0.082$.
The full width at half maximum is thus $(y_1-y_2) \, a = 4.603 \, a$.
For comparison, the FWHM of the exponential itself is 
$\log2 \; a = 0.693 \, a$.
The ratio between the brightness at the center of the sphere and at
maximum is $\mathcal{B}(0)/\mathcal{B}_{\mathrm{max}} = \sqrt{a/2} / g_0
= 1.307 \sqrt{a}$.
The explicit (non-integral) approximation proposed by \citet{bv04}
is less accurate in representing the peak, but correctly
represents the profile toward the center of the sphere 
(Eq. \ref{expproj} doesn't).
Note that the difference with the value 7 $a$ given for the width
by \citet{bv04}
is mostly due to a different definition (they consider the width
at 1/e of the maximum).


\end{document}